\documentclass[aip,rsi,reprint]{revtex4-1} 
\usepackage{graphicx}
\usepackage{booktabs}
\usepackage{multirow}
\usepackage{color}

\begin{document}


\title{A Large-Diameter Hollow-Shaft Cryogenic Motor Based on a
  Superconducting Magnetic Bearing for Millimeter-Wave Polarimetry}


\author{B.~R.~Johnson}
\email[]{bradley.johnson@columbia.edu}
\affiliation{Department of Physics, Columbia University, New York, NY 10027, USA}

\author{F.~Columbro} \affiliation{Dipartimento di Fisica, Universit\`a
  di Roma La Sapienza, 00185 Roma, Italy}

\author{D.~Araujo}
\affiliation{Department of Physics, Columbia University, New York, NY 10027, USA}

\author{M.~Limon}
\affiliation{Department of Physics, Columbia University, New York, NY 10027, USA}

\author{B.~Smiley}
\affiliation{Department of Physics, Columbia University, New York, NY 10027, USA}

\author{G.~Jones}
\affiliation{Department of Physics, Columbia University, New York, NY 10027, USA}

\author{B.~Reichborn-Kjennerud}
\affiliation{Department of Physics, Columbia University, New York, NY 10027, USA}

\author{A.~Miller} \affiliation{Department of Physics and Astronomy,
  University of Southern California, Los Angeles, CA 90089, USA}

\author{S.~Gupta}
\affiliation{Department of Physics, Columbia University, New York, NY 10027, USA}


\date{\today}


\begin{abstract}
In this paper we present the design and measured performance of a
novel cryogenic motor based on a superconducting magnetic bearing
(SMB).
The motor is tailored for use in millimeter-wave half-wave plate (HWP)
polarimeters, where a HWP is rapidly rotated in front of a
polarization analyzer or polarization-sensitive detector.
This polarimetry technique is commonly used in cosmic microwave
background (CMB) polarization studies.
The SMB we use is composed of fourteen yttrium barium copper oxide
(YBCO) disks and a contiguous neodymium iron boron (NdFeB) ring
magnet.
The motor is a hollow-shaft motor because the HWP is ultimately
installed in the rotor.
The motor presented here has a 100~mm diameter rotor aperture.
However, the design can be scaled up to rotor aperture diameters of
approximately 500~mm.
Our motor system is composed of four primary subsystems: (i) the rotor
assembly, which includes the NdFeB ring magnet, (ii) the stator
assembly, which includes the YBCO disks, (iii) an incremental encoder,
and (iv) the drive electronics.
While the YBCO is cooling through its superconducting transition, the
rotor is held above the stator by a novel hold and release mechanism
(HRM).
The encoder subsystem consists of a custom-built encoder disk read out
by two fiber optic readout sensors.
For the demonstration described in this paper, we ran the motor at
50~K and tested rotation frequencies up to approximately 10~Hz.
The feedback system was able to stabilize the the rotation speed to
approximately 0.4\%, and the measured rotor orientation angle
uncertainty is less than 0.15~deg.
Lower temperature operation will require additional development
activities, which we will discuss.
\end{abstract}




\maketitle 


\begin{figure*}[t]
\centering
\includegraphics[width=\textwidth]{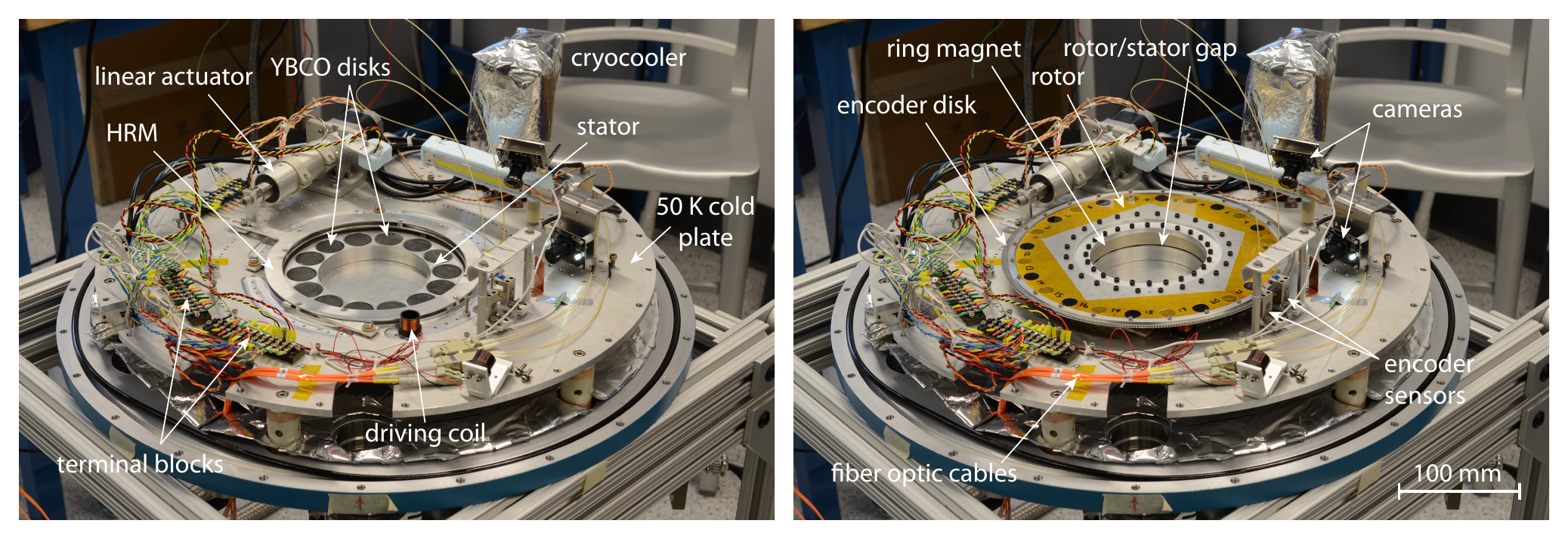}
\caption{
Photographs of the motor system installed on the 50~K cold plate of
the cryostat.
In the photograph on the left, the rotor is removed exposing the YBCO
disks and the hold-and-release mechanism (HRM).
In the photograph on the right, the HRM is engaged, holding the rotor
in place above the stator.
The 2~mm gap between the rotor and the stator is visible.
The rotor and HRM are shown in more detail in
Figure~\ref{fig:hrm_and_rotor}.
Excluding the linear actuator, the overall height of the motor is
38~mm, and the fully assembled rotor and stator fit inside a circular
footprint approximately 280~mm in diameter.
The vacuum shell of the cryostat and the 50~K radiation shield were
removed for these photographs, so the motor elements are visible.
}
\label{fig:overview_figure} 
\end{figure*}


\section{Introduction}
\label{sec:introduction}

Half-wave plate (HWP) polarimetry is a technique that is now commonly
used for cosmic microwave background (CMB) polarization
studies\cite{ebex_2017, takakura_2017, essinger-hileman_2016,
  bryan_2016, hill_2016, kusaka_2014, johnson_2007}.
With this technique a HWP is rotated in the telescope beam in front of
a polarization analyzer.
The linearly polarized component of the incoming radiation is rotated
by the HWP, ultimately producing a modulated signal in the emerging
detector data stream at four times the rotation frequency of the
HWP\cite{johnson_2004}.
If the HWP is rapidly rotated, then this approach allows the observer
to move the polarization signals away from problematic noise features
in the detector bandwidth, such as low-frequency $1/f$ noise, thereby
maximizing the sensitivity of the instrument and possibly enabling
observations on larger angular scales\cite{brown_2009}.
The approach can also be used to help mitigate the effect of some
kinds of instrument-induced systematic errors\cite{odea_2007}, but for
this paper we are primarily interested in the technical aspects of
rapid HWP rotation and ways to produce high-frequency signal
modulation.

For CMB polarization studies there are three primary performance
requirements to consider for the HWP rotation system.
First, thermal emission from the optical elements in the polarimeter
-- such as the HWP -- must be minimized to maximize instrument
sensitivity because the noise from state-of-the-art millimeter-wave
detector systems is commonly limited by the random arrival of photons.
Since the CMB is a 2.7~K blackbody, the brightness temperature of any
thermal emission in the instrument ideally should be suppressed to
approximately this level or below.
Second, because the detector systems are photon-noise limited, large
arrays of independent detectors are required to average down the
instrument noise.
This requirement has created the need for large-throughput (A$\Omega$
product) telescopes, which ultimately means the rotating HWP must have
a large diameter (between approximately 200~mm and 500~mm).
Third, to accurately reconstruct the polarization patterns on the sky,
the HWP orientation must be measured to approximately 0.2~deg or
better\cite{odea_2007}.
These three requirements together create the need for a
speed-variable, large-diameter, hollow-shaft rotator that operates at
cryogenic temperatures.
The orientation of the rotator must be precisely measured, and the
cryogenic parts of the system cannot generate appreciable amounts of
heat from mechanisms like stick-slip friction.

To solve this problem, the balloon-borne EBEX experiment
developed,\cite{hanany_2003, matsumura_2005a, matsumura_2005b,
  matsumura_2006} built, and deployed\cite{klein_2011} a HWP rotator
based on a superconducting magnetic bearing (SMB).
In the EBEX SMB, a permanent neodymium iron boron (NdFeB) ring magnet
assembly was levitated above a matching ring of yttrium barium copper
oxide (YBCO) via field cooling.
SMBs based on high-temperature superconductors are attractive for this
application because flux pinning allows any-orientation operation
after the bearing rotor is levitating\cite{moon_1994}.
The EBEX HWP, which was mounted inside the rotor of the SMB, was
ultimately rotated by a motor mounted outside the cryostat.
The driveshaft of this motor passed through the cryostat shell using a
rotary vacuum feedthrough and turned a small-diameter pulley wheel
inside the cryostat that was coupled to the rotor of the SMB via a
Kevlar belt.

Many groups around the world are developing rotators for HWP
polarimetry using different approaches\cite{matsumura_2016a,
  matsumura_2016b, hill_2016, bryan_2016}.
In this paper we present the design and measured performance of one
novel approach: a hollow-shaft, cryogenic motor based on an SMB.
We describe the motor hardware in Section~\ref{sec:methods}, and the
characterization measurement results in Section~\ref{sec:results}.
The success of this proof-of-concept prototype motor justifies future
development activities, which are discussed in
Section~\ref{sec:discussion}.


\section{Methods}
\label{sec:methods}

Our motor system is composed of four primary subsystems: (i) the rotor
assembly, (ii) the stator assembly, which includes the YBCO, (iii) the
encoder, and (iv) the drive electronics.
An overview photograph of the motor system is shown in
Figure~\ref{fig:overview_figure}, and the key elements in each
subsystem are described in Sections~\ref{sec:cryostat} to
\ref{sec:drive_system}.

Our design builds from the EBEX design and makes three critical
advancements.
First, our hold-and-release mechanism (HRM), which is described in
Section~\ref{sec:hold_and_release_mechanism}, could prove to be more
mechanically robust than the EBEX design that used spring loaded
``grippers'' positioned by a linear actuator and a Kevlar-string
pulley system.
We show in this paper that our HRM functions cryogenically as
designed, and although our prototype hollow-shaft diameter is just
100~mm, our HRM design is scalable and can ultimately accommodate HWP
diameters between 100~mm and 500~mm or more\cite{araujo_2014,
  johnson_2014}.
Second, the rotor can be driven with a single electromagnet coil pair,
so the Kevlar belt is not needed in our approach.
This design change removes a thermal load on the rotor, which should
help minimize the HWP temperature.
Third, our optical encoder design, which is based on optical fibers
(Section~\ref{sec:encoder}), moves the required light sources and
photodetectors outside the cryostat, which removes a heat load on the
cryogenics, and our quadrature readout approach provides additional
rotation direction information.

Since this motor is the product of a proof-of-concept study, two
additional choices constrained our design.
First, for convenience at this early development stage, we targeted
50~K operation, noting that lower-temperature operation may be more
desirable in the future.
Second, after considering segmented magnets and magnet tiles for the
ring on the rotor, we decided that a contiguous magnet was the most
promising in terms of SMB performance.
Therefore, the scale of our prototype motor was set by the largest
commercially available off-the-shelf NdFeB ring magnet we could find.
We then tailored the HRM to the size of this magnet.
Larger magnets can be fabricated as custom parts, so our approach is
scalable.


\begin{figure*}[t]
\includegraphics[width=\textwidth]{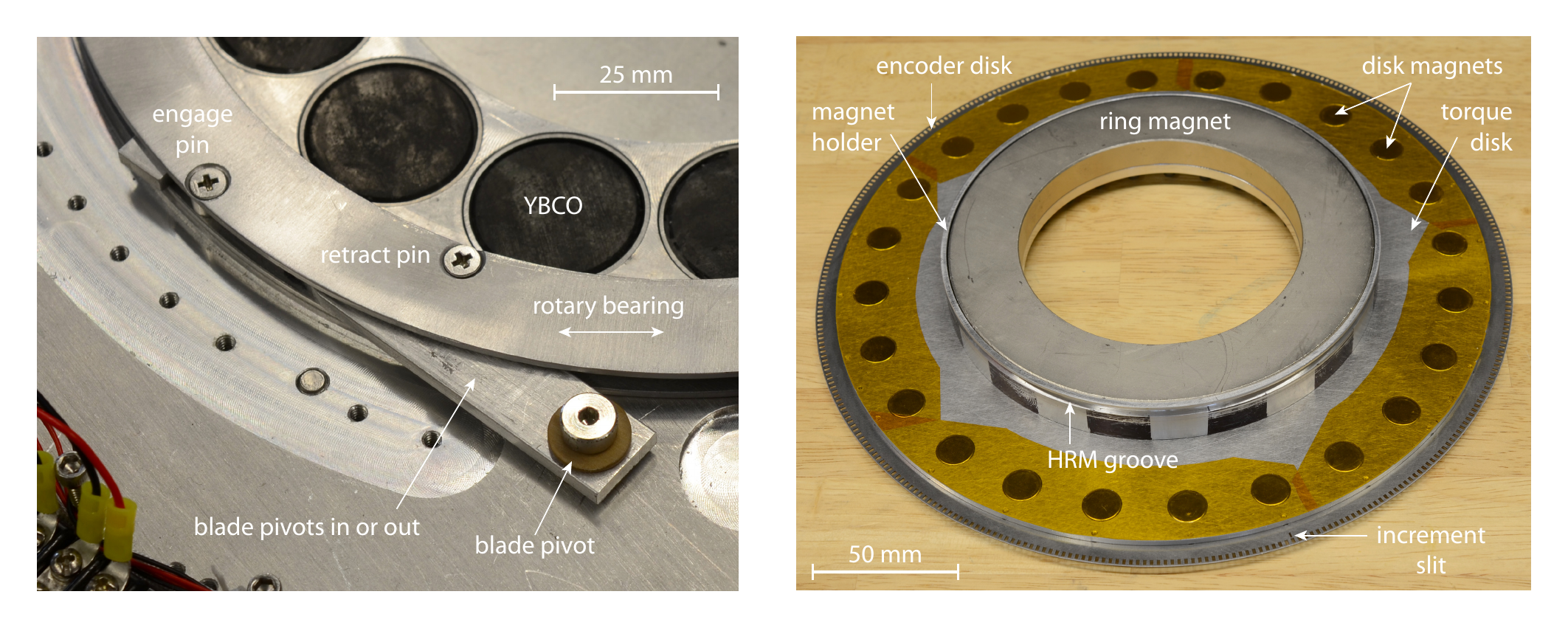}
\caption{
\textbf{Left:} A photograph of one of the three hold-and-release
mechanism blades.
The linear actuator shown in Figure~\ref{fig:overview_figure} is used
to move the rotary bearing.
If the rotary bearing moves clockwise (in the photograph), then the
retract pin pivots the blade away from the rotor.
If the rotary bearing moves counterclockwise, then the engage pin
pivots the blade toward the rotor and ultimately into its HRM groove.
\textbf{Right:} A photograph of the reverse side of the rotor
(relative to the view shown in Figure~\ref{fig:overview_figure}).
The rotor builds from an aluminum ring-magnet holder that has the HRM
groove.
A separate aluminum ``torque disk'' serves as a mount for the disk
magnets used in the drive system and a mount for the encoder ring.
The ring-magnet holder and the torque disk are assembled with steel
screws, and the ring magnet is held in place using the attractive
force between the ring magnet and these steel screws; no epoxy was
used.
The steel screws (black) are visible in
Figure~\ref{fig:overview_figure}.
After the YBCO in the stator cools below its superconducting
transition temperature, the linear actuator is retracted, opening the
HRM, releasing the rotor and allowing it to levitate freely.
}
\label{fig:hrm_and_rotor}
\end{figure*}


\subsection{Cryostat}
\label{sec:cryostat}

The motor is mounted inside a custom-built aluminum
cryostat\footnote{Precision Cryogenics Systems,
  Model~\#:~PCS~21.0/14.0} on an aluminum cold plate 530~mm in
diameter and 6.4~mm thick.
The cold plate is cooled to 50~K by a two-stage Gifford-McMahon (GM)
cryocooler\footnote{CTI-Cryogenics, Cryodyne Refrigeration System,
  Model 350 (two-stage configuration)}.
For this work, only the 50~K stage of the GM cooler is used; the
second, 20~K stage is unloaded, which gives the 50~K stage more
cooling power.
A cylindrical radiation shield mounted to the cold plate surrounds the
motor to reject 300~K radiation from the vacuum shell of the cryostat.
The radiation shield is 530~mm in diameter and 350~mm tall.
The temperature of the system is monitored with four-wire platinum
resistor temperature sensors\footnote{Lake Shore Cryotronics, Inc.,
  Part~\#:~PT-103-AM-2S and Part~\#:~PT-103-AM}.
One sensor is mounted directly on the cold head of the GM cryocooler,
while the other three are mounted on the cold plate near the YBCO
disks on the stator.
The thermometers are read out using commercially available
electronics\footnote{Lake Shore Cryotronics, Inc., Temperature
  Monitor, Model 218}.

The vacuum shell on the cryostat contains four KF50 ports.
The first port is used for vacuum pumping\footnote{Pfeiffer D-35164
  Asslar turbo pump backed by a BOC Edwards XDS10 dry scroll vacuum
  pump}.
The second port hosts a USB 2.0 vacuum feedthrough\footnote{MDC Vacuum
  Products, USB K150 DN40 KF FLG MT, Part~\#: 9173001} that is used to
provide power and signal I/O wiring for two USB cameras (see
Section~\ref{sec:visual_inspection}).
The third port hosts a custom-built vacuum feedthrough for the optical
fibers used in the encoder system\footnote{Douglas Electrical
  Components, custom fiber optic vacuum feedthrough}.
The fourth port hosts a vacuum feedthrough for a 41-pin
connector\footnote{Ceramtek, 41-pin NW50KF vacuum feedthrough, and
  Amphenol Industrial Operations, 41-pin connector,
  Part~\#:~PT06SE-20-41S(SR)}, which is used for all additional
wiring.
The electrical current driving the motor is appreciable so we use
24~AWG copper wire for the two drive-coil wires (see
Section~\ref{sec:drive_system}).
Otherwise, to minimize the thermal loading on the 50~K stage we use
32~AWG manganin 290 wire\footnote{California Fine Wire Company}.
Terminal blocks mounted on the cold plate (see
Figure~\ref{fig:overview_figure}) serve as a thermal intercept for the
wiring and a convenient electrical interconnect.


\begin{figure*}[t]
\centering
\includegraphics[width=0.95\textwidth]{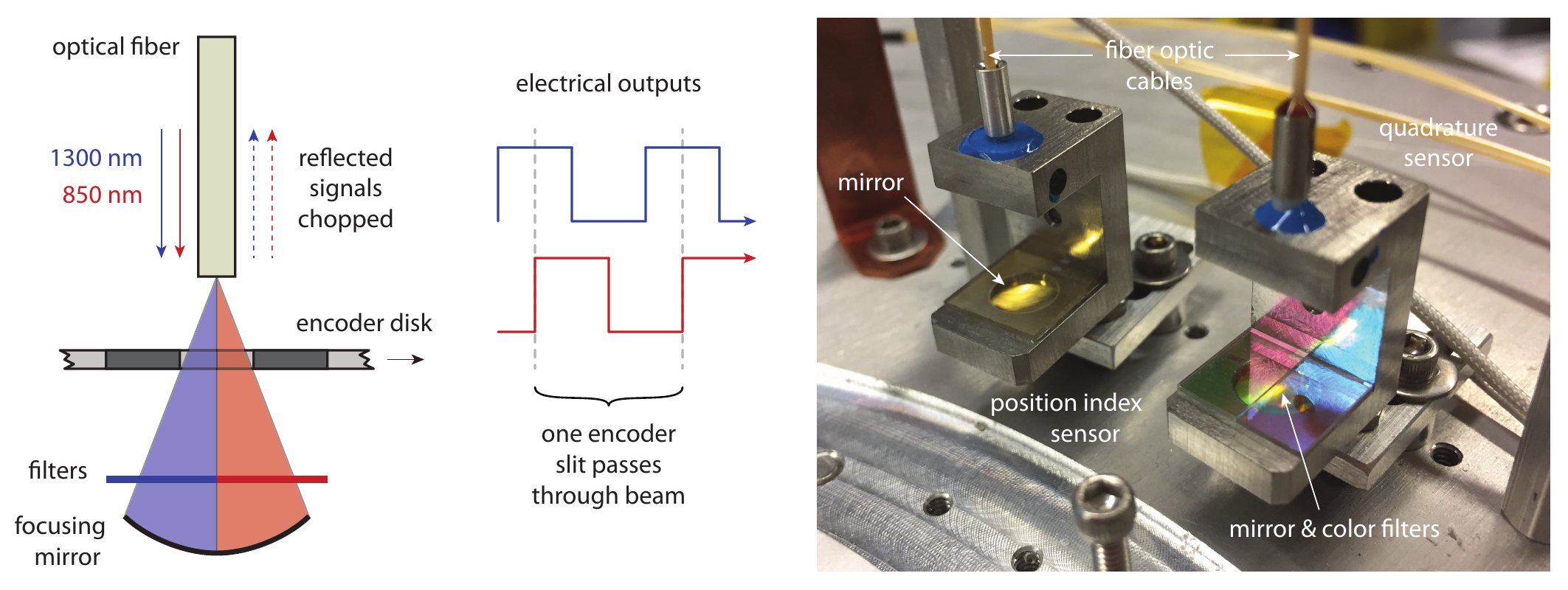}
\caption{
\textbf{Left:} Schematic illustrating how the quadrature encoder
sensor works.
Two wavelengths of light are carried into the cryostat on a single
optical fiber.
The light emitted from the end of the optical fiber is filtered and
then refocused back into the fiber with a mirror.
The filters are arranged so the left half of the reflected beam passes
1300~nm, while the right half passes 850~nm.
As the incremental encoder disk passes by the beam is chopped on and off.
However, because one filter precedes the other, the two wavelengths
are chopped out of phase, and this quarter-wave phase offset indicates
the rotation direction.
The encoder readout transforms the optical signals into a quadrature
electrical outputs, which can be used in the feedback loop and to
determine the orientation angle of the rotor.
\textbf{Right:} A photograph of the two encoder sensors mounted in the
cryostat on the 50~K cold plate.
For clarity, the encoder disk is not shown.
Note that the position index sensor detects a single increment slit on
a separate track of the encoder disk each rotation (see the right
panel of Figure~\ref{fig:hrm_and_rotor}).
Therefore, only a single wavelength of light (850~nm) is used for this
sensor.
}
\label{fig:encoder}
\end{figure*}


\subsection{Superconducting Magnetic Bearing}
\label{sec:smb}

The stator of the superconducting magnetic bearing consists of
fourteen YBCO disks\footnote{CAN Superconductors, s.r.o., Czech
  Republic} that are arranged in a ring.
This ring can be easily seen in the left panel of
Figure~\ref{fig:overview_figure}.
Each YBCO disk is 13~mm tall, 25~mm in diameter and is specified to
generate 60~N of levitation force at 77~K assuming the magnet/YBCO
separation distance is 9~mm and the permanent magnet has a 0.5~T field
at its surface.
YBCO is a brittle material, so before assembly, each disk was first
mounted in a thin-walled cylindrical aluminum cup with
epoxy\footnote{Stycast 2850 FT}.
Each cup is open on the top and has tapped mounting holes on the
bottom.
Each disk/cup assembly was then screwed into a recess in a
custom-built aluminum holder on the stator.
The holder defined the ring shape.
This mounting approach yields good thermal contact between the YBCO
and the 50~K cold plate, while at the same time, it allows disassembly
and disk replacement.

The rotor of the SMB is an N42 grade NdFeB ring
magnet\footnote{Applied Magnets, Model~\#:~NR025} with a nominal pull
force of 930~N and a residual flux density of 1.3~T.
The ring magnet is 13~mm thick, it has a 150~mm outer diameter, a
100~mm inner diameter, and a mass of 0.88~kg.
To provide durability and protection against corrosion, the magnet was
coated with a nickel-copper-nickel trilayer.
During cryogenic operation, the rotor levitated above the stator with
a rotor/stator separation distance of approximately 2~mm.


\subsection{Hold and Release Mechanism}
\label{sec:hold_and_release_mechanism}

While the YBCO cools, the rotor is held above the stator by the hold
and release mechanism shown in Figure~\ref{fig:hrm_and_rotor}.
The HRM design was inspired by an iris diaphragm.
It is composed of (i) a large-diameter, custom-built rotary bearing
that encircles the YBCO ring on the stator, (ii) three pivoting
blades, (iii) engage and retract pins on the rotary bearing that pivot
the blades in and out, and (iv) a linear
actuator\footnote{UltraMotion, Part~\#:~D-A.083-HT17-2-2NO-BR/4-CRYO}
that moves the rotary bearing by pushing or pulling tangentially on
its edge.
During operation, as the linear actuator moves, the rotary bearing
turns, and the blades move in or out.
When the motor is at room temperature the linear actuator is extended,
and the HRM blades slot into a groove on the edge of the rotor,
thereby holding the rotor fixed above the stator both vertically and
laterally.
After the YBCO is cooled below the superconducting transition
temperature (93~K) the linear actuator is retracted, causing the HRM
to remove the blades, release the rotor, and allow the rotor to
levitate freely above the stator.
The linear actuator, which was designed to function at cryogenic
temperatures, is powered by a two-phase, high-torque, hybrid stepper
motor\footnote{Applied Motion, Part~\#:~HT17-75} with a resolution of
95~steps per millimeter and 200~steps per revolution.
The motor driver\footnote{Applied Motion, Part~\#:~ST5-Si-NN} was
pre-programmed to either extend or retract the actuator by 25~mm upon
activation.
Limit switches provided with the stepper motor prevent the linear
actuator from overextending and damaging the HRM or the rotor.


\subsection{Encoder}
\label{sec:encoder}

The encoder system consists of a custom-built incremental encoder disk
read out by two optical readout sensors\footnote{Micronor Inc.}.
The right panel of Figure~\ref{fig:hrm_and_rotor} shows a photograph
of the encoder disk, and Figure~\ref{fig:encoder} illustrates the
operation of the encoder readout system.

The encoder disk, which serves as a photo interrupter, has 360 equally
spaced slits around its perimeter, and it is mounted to the edge of
the ``torque disk'' on the rotor (see Figure~\ref{fig:hrm_and_rotor}
and Section~\ref{sec:drive_system}).
The encoder disk was laser cut from 400~$\mu$m thick stainless steel
sheet metal.
Each slit is approximately 2~mm long and 1.11~mm wide at its widest
point.
The outer diameter of the encoder disk is 254~mm, while the inner
diameter is 229~mm.

Two wavelengths of light (1300 and 850~nm) are generated by sources
outside the cryostat\footnote{Micronor Inc., Fiber Optic Encoder
  System, Model~\#:~MR320} and fed into the cryostat on a optical
fiber that passes through the aforementioned custom-built, fiber-optic
vacuum feedthrough.
The total diameter of each optical fiber is 125~$\mu$m, and the core
diameter is 62.5~$\mu$m.
Inside the cryostat, the end of the optical fiber is mounted
perpendicular to the encoder disk near the slits using custom-built
titanium mounts.
Light emitted from the end of the optical fiber is filtered and then
refocused back into the fiber with a spherical mirror.
This reflected light is detected outside the cryostat with
photodetectors.
The filters, which are mounted in front of the focusing mirror, are
arranged so the left half of the reflected beam passes only 1300~nm
light, while the right half passes only 850~nm light.
During operation, the emerging beam passes through, or is interrupted
by, the slits in the encoder disk.
However, because one filter precedes the other, the two wavelengths
are chopped out of phase, and this quarter-wave phase offset indicates
the rotation direction (see left panel of Figure~\ref{fig:encoder}).
The encoder readout transforms the optical signals into quadrature
electrical square wave signals, which are used in the feedback loop to
determine the orientation angle, direction, and angular speed of the
rotor.
Note that inside the optical fiber, light is traveling in both directions.
The in-bound light beams are continuous, while the out-bound beams are
chopped.

A second optical sensor detects a single index position slit on a
separate inner track of the encoder disk using an 850~nm light
beam\footnote{Micronor, Inc., Fiber Optic U-Beam Controller Module,
  Model~\#:~MR382-1-1}.
This index position slit resets an internal counter in the feedback
system each revolution, and it is used to ensure that the orientation
angle of the rotor is reset to zero after each complete rotation of
the encoder disk (see right panel of Figure~\ref{fig:hrm_and_rotor}).


\begin{figure}[t]
\centering
\includegraphics[width=\columnwidth]{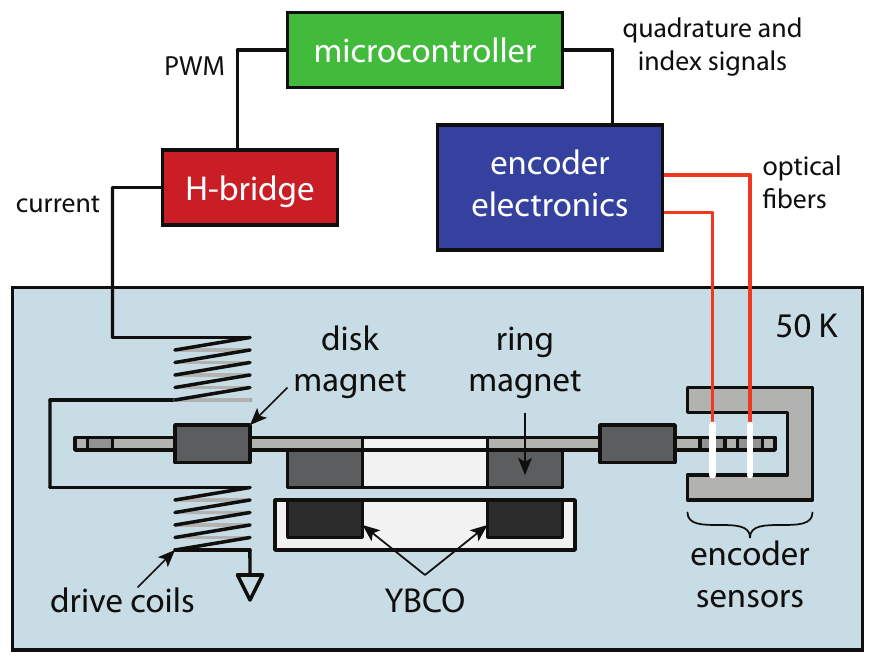}
\caption{
A schematic showing the feedback loop of the motor control system.
The user first inputs the target rotation frequency into the software
on the microcontroller.
The microcontroller then sends a PWM signal to the H-bridge, which
controls the current in the drive coils.
The primary motor elements mounted at 50~K are shown in cross section.
The magnetic field produced by the coils interacts with the ring of
permanent disk magnets that are mounted on the rotor (see
Figure~\ref{fig:hrm_and_rotor}).
As the rotor turns, the optical encoder generates the index position
pulse and the quadrature square-wave signals that are shown in
Figure~\ref{fig:encoder}.
These signals are fed back into the microcontroller, which uses them
to compute the angular position and speed of the rotor.
The output PWM signal is automatically adjusted by the microcontroller
if the measured rotation frequency is different from the input target
rotation frequency.
}
\label{fig:feedback_loop}
\end{figure}


\begin{figure*}[t]
\centering
\includegraphics[width=0.95\textwidth]{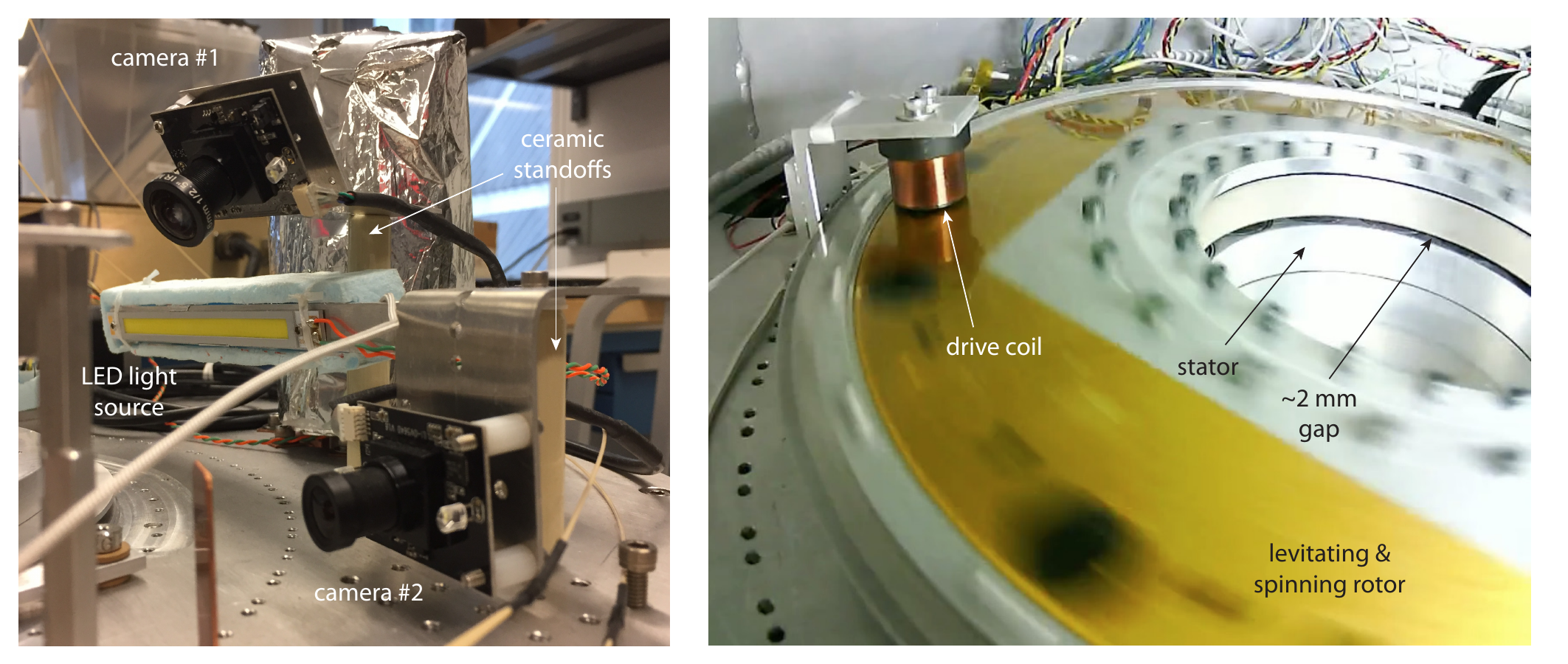}
\caption{
\textbf{Left:} A photograph of the two cameras that are mounted inside
the cryostat.
These cameras allowed visual monitoring during 50~K operation.
\textbf{Right:} One frame from a movie that was recorded using
camera~\#1.
The rotor is levitating and rotating.
}
\label{fig:cameras} 
\end{figure*}


\subsection{Drive System}
\label{sec:drive_system}

The drive system for the motor uses a proportional-integral (PI)
feedback loop composed of a microprocessor, an
H-bridge\footnote{TrossenRobotics.com, L298 Dual H-Bridge Motor
  Driver, Part~\#: SS-MOT103B1M}, two hollow-core electromagnet
coils\footnote{Moticont, Hollow Core Linear Voice Coil,
  Part~\#:~HVCM-019-016-003-02-COIL}, and an optical encoder.
The electromagnet coils are 10.8~mm long, have an inner diameter of
13.4~mm, an outer diameter of 16.0~mm, and are composed of
approximately 120 turns of insulated 30~AWG copper wire.
A schematic of the feedback loop is shown in
Figure~\ref{fig:feedback_loop}.
We chose to use the Arduino~2 microcontroller because it has a fast
84~MHz clock and pulse width modulation (PWM) output pins.
Rotation frequencies between 2.5 and 25~Hz (150 and 1500 RPM) are
useful because the $1/f$ knee of low-frequency noise in
millimeter-wave observations is commonly 10~Hz or
below\cite{takakura_2017} and millimeter-wave detector systems can
have up to approximately 100~Hz or more of usable
bandwidth\cite{mccarrick_2014}.
For the demonstration described in this paper, we tested rotation
frequencies up to approximately 10~Hz.

During operation, the rotation direction and the target rotation
frequency of the rotor are defined by the user as input parameters in
the software on the microcontroller.
The microcontroller then sends a PWM signal to a standard H-bridge
circuit that controls both the magnitude and polarity of the current
through the coils.
The time-varying magnetic field produced by the coils interacts with
the ring of 24 permanent disk magnets that are mounted on the torque
disk on the rotor (see Figure~\ref{fig:hrm_and_rotor}).
The disk magnets\footnote{K\&J Magnetics, Inc., Part~\#:~D82} are
13~mm in diameter, 3.2~mm thick, made from grade N42
nickel-copper-nickel plated NdFeB, and arranged so any two neighboring
magnets in the ring have opposite polarity.
The drive coils are mounted approximately 2~mm from the disk magnets.
The torque disk is used to separate the disk magnets from the ring
magnet and the YBCO.
This ensures that flux from the disk magnets is not pinned in the
YBCO.
It also increases the torque applied by the drive coil pair because
the force is applied at a larger radius.

When the rotor is initially at rest, strong current pulses are sent to
the coils to start the rotor turning.
When the position index sensor detects a pulse for the first time the
current modulation begins.
During normal operation, the optical encoder generates the index
position pulse and the quadrature square-wave signals that are shown
in Figure~\ref{fig:encoder}.
These signals are fed back into the microcontroller, which uses them
to compute the angular position and speed of the rotor.
The angular speed of the rotor is the feedback signal for the PI loop.
The rotor speed and orientation measurements are also output and
stored on a computer as time-ordered data.


\section{Results}
\label{sec:results}


\subsection{Visual Inspection}
\label{sec:visual_inspection}

Our cryostat does not have a window.
Therefore, we installed two USB~2.0 cameras\footnote{Leopard Imaging,
  5M HD USB Camera, Part~\#:~LI-OV5640-USB-72} inside the cryostat to
watch the motor turn and to observe the HRM during operation (see
Figure~\ref{fig:cameras}).
The cameras we selected use the 1/4 inch OmniVision OV5640 CMOS
sensor, which has 1.4~$\mu$m square pixels and a resolution of $2592
\times 1944$.
The camera modules use threaded M12 lens mounts, and can deliver
either still photographs or video at up to 30~fps.
With camera~\#1 we use an F/2.0 lens with a 2.6~mm focal length to
observe the system as a whole from the top of the motor.
Camera~\#2 uses an F/2.6 lens with a 2.8~mm focal length to monitor
any lateral movement in the rotor from the side.
The camera power and signal I/O were fed into the cryostat through the
aforementioned USB 2.0 vacuum feedthrough (see
Section~\ref{sec:cryostat}).

These cameras are not nominally rated for cryogenic operation.
Therefore, we took precautions to make sure they would function
properly in our system.
We first mounted the bare camera modules on custom aluminum interface
plates with standoffs to prevent electrical shorts on the back side of
the printed circuit boards.
We then mounted these assemblies on ceramic
standoffs\footnote{Isolantite Manufacturing Company, Inc.} to reduce
the thermal conductance between the cameras and the 50~K cold plate.
In the vacuum environment inside the cryostat, there is no convective
heat transfer and radiative heat transfer is negligible.
Therefore in equilibrium, waste electrical heat flowing into the
camera module is equal to the heat flowing out of the camera module
through the ceramic standoff.
For our thermal design, the associated equilibrium temperature falls
within the operating range of the camera.
The camera modules included white LEDs for subject illumination, but
we found them to be too dim.
Therefore, we added a 120~mm long white linear LED\footnote{Vollong,
  3W White High Power Linear COB LED, Part~\#:~VL-H03W5500380D12} that
has a nominal luminous intensity of 36,600~mcd.
The linear LED was not rated for cryogenic operation, so we used a
similar thermal circuit to keep it sufficiently warm.

The visibility the cameras provided was particularly useful during the
initial stages of the project when the HRM was being developed.
As an example, without the cameras it was difficult to know if the
rotor released properly when the linear actuator retracted the HRM
blades.
We discovered early on that the rotor unexpectedly drops approximately
1~mm after release, decreasing the size of the rotor/stator gap.
This drop caused a mechanical conflict between early versions of the
the rotor and the stator, and this mechanical conflict initially
prevented the rotor from turning.
The problem was observed through a real-time video stream.
Therefore, the cameras helped hone the HRM design by revealing flaws
that otherwise would not have been visible.


\begin{figure}[h]
\centering
\includegraphics[width=0.48\textwidth]{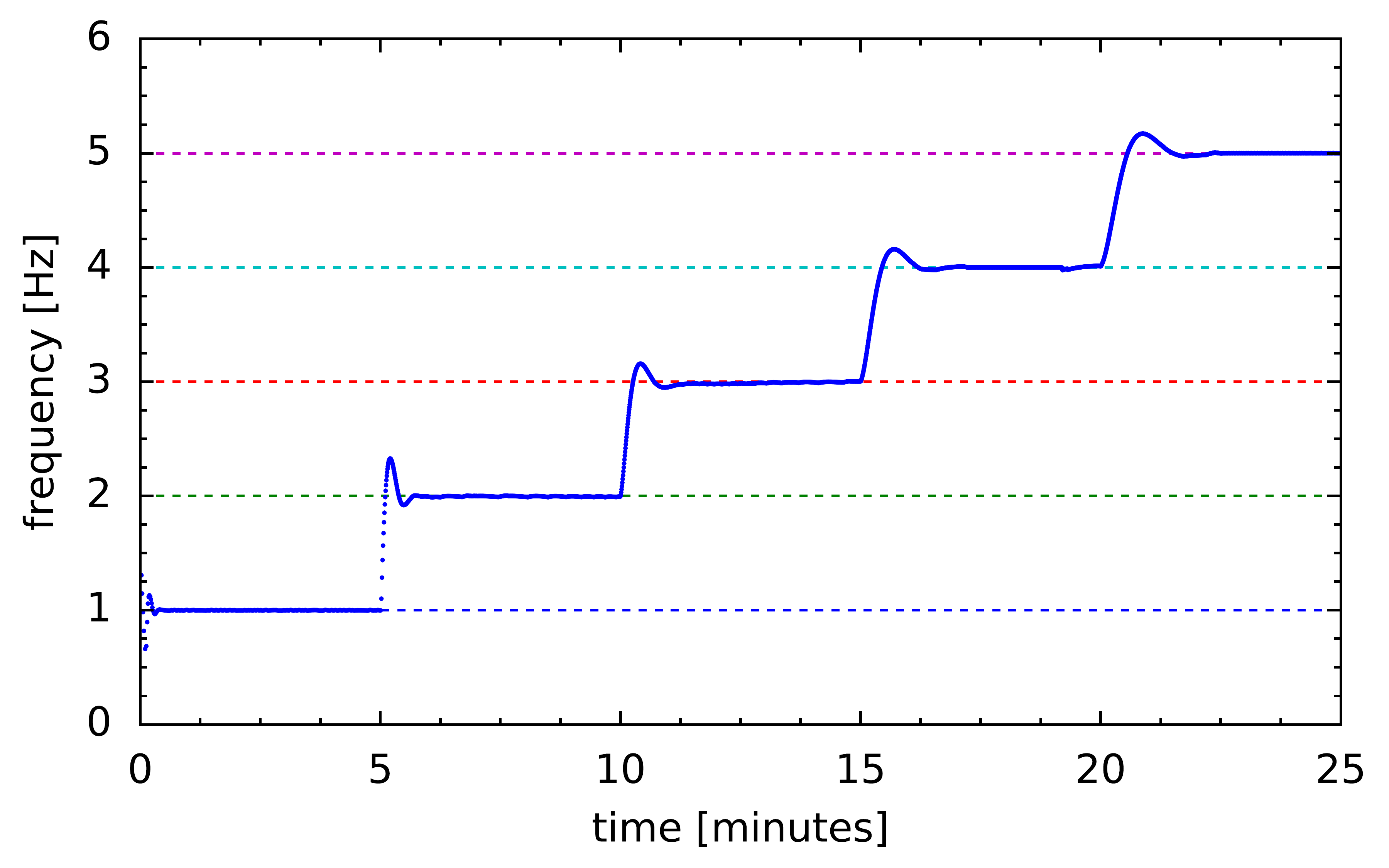}
\caption{
Rotation frequency as a function of time.
The target rotation frequency was changed every 5 minutes.
The horizontal dashed lines indicated the target rotation frequencies.
After settling, the feedback system successfully maintained the
desired rotation frequency.
Though the plot shows data for rotation frequencies up to 5~Hz, we
tested rotation frequencies up to approximately 10~Hz.
}
\label{fig:signal_vs_time}
\end{figure}


\begin{figure*}[t]
\centering
\includegraphics[width=0.49\textwidth]{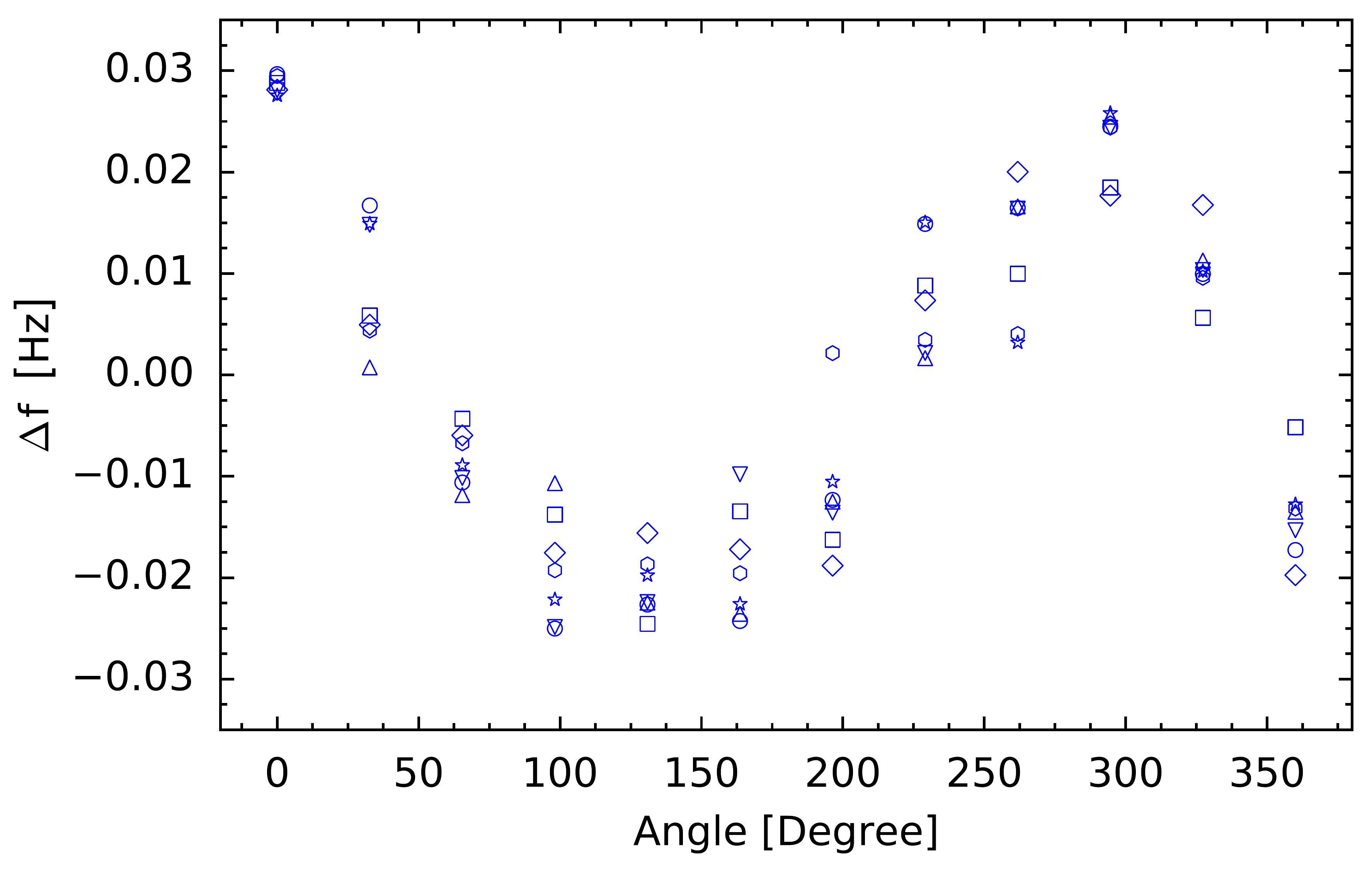}
\includegraphics[width=0.49\textwidth]{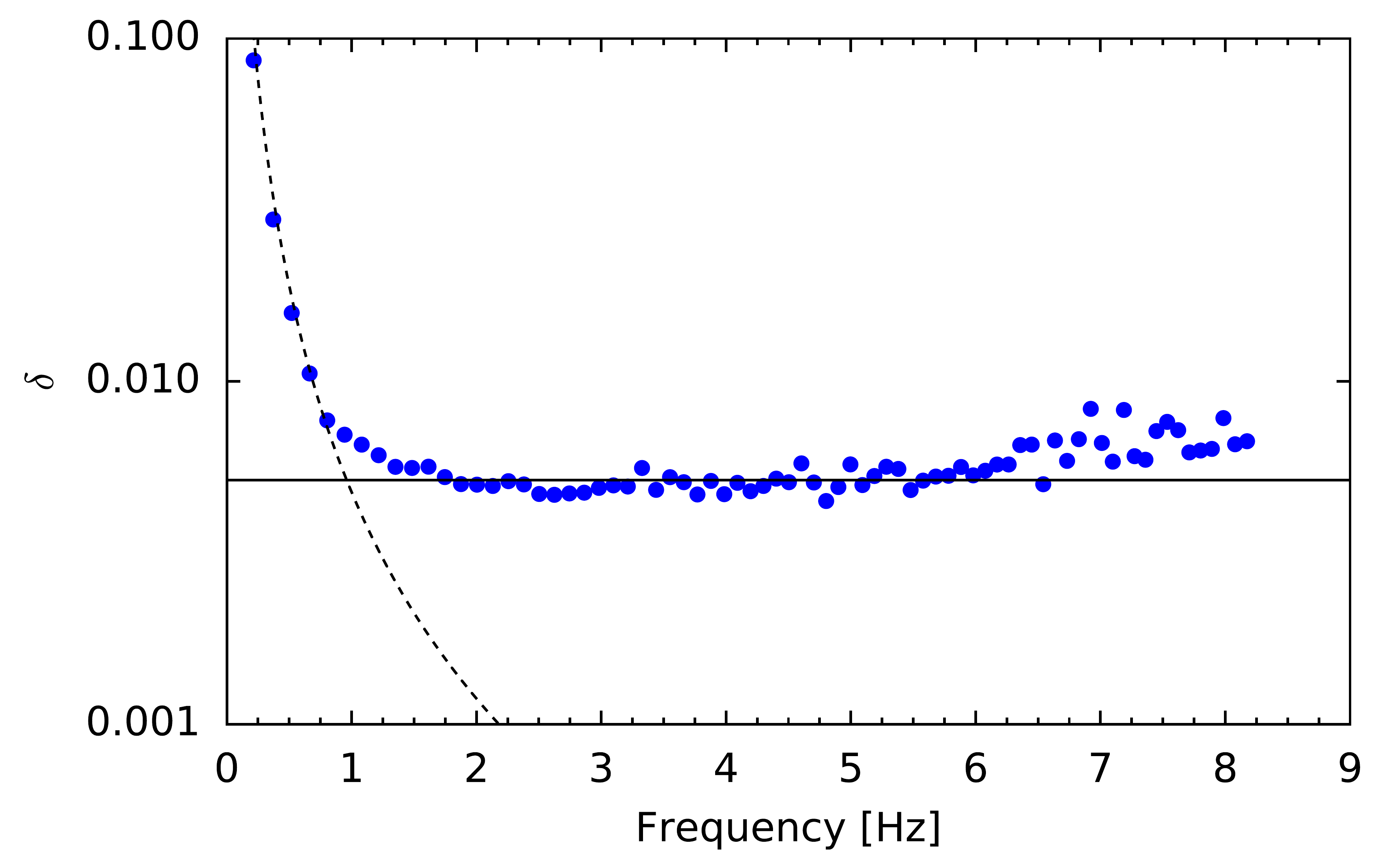}
\caption{
\textbf{Left:} Variation in the rotation frequency as a function of
rotor orientation angle for seven periods.
For this measurement, the rotor was rotating at 7~Hz (see
Section~\ref{sec:speed_and_stability}).
The $\Delta f_0$ residual, which is approximately sinusoidal, is
apparent (see Equation~\ref{eq:Delta_f_0}).
\textbf{Right:} Fractional change in the rotation frequency as a
function of rotation frequency.
The dashed curve is Equation~\ref{eq:approx} fit to the data below
1~Hz, while the solid curve shows the average value of $\delta$
between 2 and 6~Hz (see Section~\ref{sec:orientation_uncertainty}).
}
\label{fig:delta} 
\end{figure*}


\subsection{Speed and Stability}
\label{sec:speed_and_stability}

When commissioning the motor, we tested the performance of the PI
feedback loop by changing the target rotation frequency in the
microcontroller every five minutes and recording the measured rotation
frequency as a function of time.
Twenty-five minutes of time-ordered data from this test are plotted in
Figure~\ref{fig:signal_vs_time}.
The data show that the feedback loop takes approximately 30 seconds to
a minute to settle, and then the rotation frequency of the rotor
stabilizes to the target rotation frequency.
We tested rotation frequencies up to approximately 10~Hz.

When analyzing the data we discovered there is some residual variation
in the angular speed after the feedback loop has stabilized.
This residual variation is approximately
\begin{equation}
\Delta f_0 = \frac{\Delta f_{pp}}{2} \, \sin ( \, 2 \pi f_0 t + \phi \, ),
\label{eq:Delta_f_0}
\end{equation}
where $\Delta f_{pp}$ is the peak-to-peak speed variation in Hz, $f_0$
is the rotation frequency in Hz, and $\phi$ is an arbitrary phase.
To clearly show this residual we plotted $\Delta f_0$ versus
orientation angle for several rotational periods in the left panel of
Figure~\ref{fig:delta}.
For this measurement, the rotation frequency was 7~Hz and $\Delta
f_{pp}$ was approximately 0.06~Hz.
Therefore, the rotation speed was stable at the level of $\pm$0.4\%.


\subsection{Orientation Uncertainty}
\label{sec:orientation_uncertainty}

Since the encoder disk has 360 slits, it is straightforward to
determine the orientation angle of the rotor to within 0.5~deg by
simply analyzing the square-wave output from the encoder electronics
with a counter.
It is possible to decrease this uncertainty by further analyzing the
rotation frequency versus time data.
Since $\dot{\theta} = 2 \pi ( f_0 + \Delta f_0 )$, we can integrate to
find $\theta$ as a function of time.
Ideally, $\theta = 2 \pi f_0 t$, but the observed residual angular
speed variation term yields an orientation angle deviation $\Delta
\theta$.
If we define the fractional change in rotation frequency as
\begin{equation}
\delta = \frac{\Delta f_{pp}}{f_0} = \frac{f_{max}-f_{min}}{f_0},
\end{equation}
then, using Equation~\ref{eq:Delta_f_0}, the maximum deviation in the
angular position of the rotor is
\begin{equation}
\Delta \theta_{max} = \frac{\delta}{2}.
\end{equation}
The right panel of Figure~\ref{fig:delta} shows measurements of
$\delta$ as a function of $f_0$.
There are three distinct regions in the plot: below 2~Hz, above 6~Hz,
and in between 2 and 6~Hz.
In the region between 2 and 6~Hz, the average value for $\delta = 5.2
\times 10^{-3}$, which means $\Delta \theta_{max}$ = 0.15~deg.
This orientation angle uncertainty is a conservative estimate, and it
meets our performance requirement (see
Section~\ref{sec:introduction}).
Above 6~Hz there is a detectable increase in $\delta$.
This increase is probably due to an observed mechanical resonance near
$\sim$12~Hz.

Below 2~Hz the data show a $1/f_0^2$ trend.
If we model the interaction between the ring magnet and the magnetic
field imprinted in the HTS as a dipole-dipole
interaction~\cite{matsumura_2006} we can define the fractional speed
variation $\delta$ with respect to the target rotation frequency $f_0$
as
\begin{equation}
\delta=\frac{\Delta f_{pp}}{f_0}=\sqrt{1+\frac{\alpha}{f_0^2}}-1
\end{equation}
where $\alpha$ is a constant related to the physical properties of the
system.
In the limit $\alpha/f_0^2\ll1$
\begin{equation}
\label{eq:approx}
\delta=\frac{\alpha}{2f_0^2},
\end{equation}
which has the observed $1/f_0^2$ dependence.
This suggests the low-frequency increase in $\delta$ is produced by
this dipole-dipole interaction.
By fitting Equation~\ref{eq:approx} to the data below 1~Hz we find
$\alpha = 9.5 \times 10^{-3}~\mbox{Hz}^{2}$.


\begin{figure}[h]
\centering
\includegraphics[width=\columnwidth]{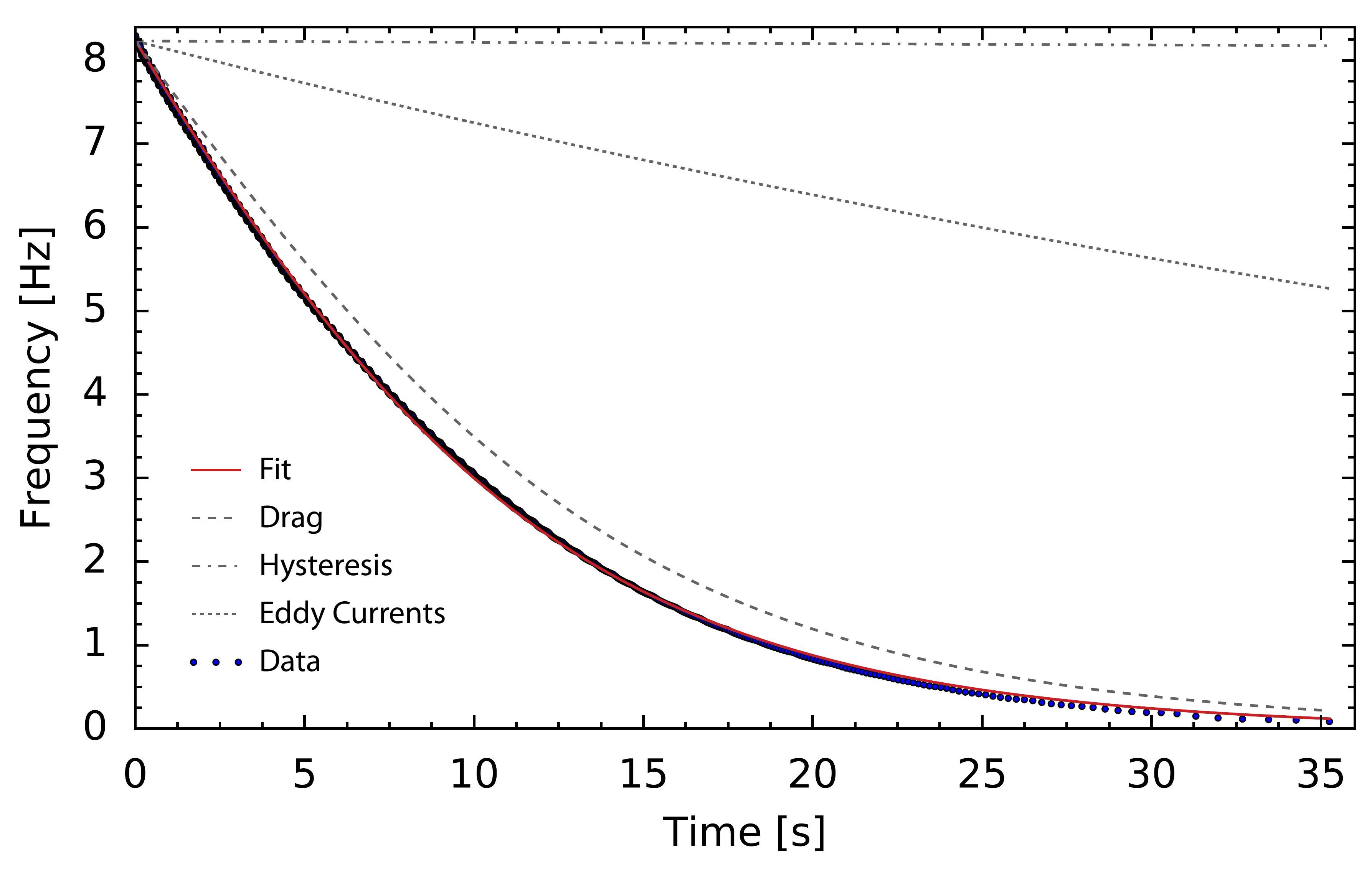}
\caption{
Spin-down test from 8.5~Hz to 0.2~Hz.
Equation~\ref{eq:eqmotion2}), which takes into account the three loss
mechanisms discussed in Section~\ref{sec:loss}, was fit to the data.
The red line is the best-fit model.
The gray dashed, dash-dot, and dotted lines represent the three terms
of model plotted individually using the best-fit parameters.
}
\label{fig:spin_down} 
\end{figure}


\subsection{Loss}
\label{sec:loss}

Although the SMB does not have stick-slip friction, other loss
mechanisms produce torque that must be overcome by the drive system.
To understand the loss mechanisms in the motor, we conducted spin-down
tests where we first turned the rotor at 8.5~Hz and then switched off
the drive system.
The unwanted torque acting on the rotor caused it to slow down and
stop after approximately 35~seconds.
Data from one spin-down test is shown in Figure~\ref{fig:spin_down}.
By fitting a theoretical model to this data, it is possible to
determine the primary loss mechanism.
Our model is composed of three loss mechanisms: hysteresis loss, eddy
current loss, and disk-magnet drag.

The drag force due to hysteresis loss~\cite{bean_1964} scales as
\begin{equation}
F_{H} \propto \frac{(\Delta B)^3}{J_c},
\label{eq:hysteresis}
\end{equation}
where $\Delta B$ is the peak-to-peak change in the magnetic field in
the superconductor, and $J_c$ is the critical current of the
superconductor.
The drag force due to eddy current loss~\cite{richards_1972} scales as
\begin{equation}
F_{EC}\propto \sigma (\Delta B)^2 f
\label{eq:eddy}
\end{equation}
where $\sigma$ is the electrical conductivity of surrounding metals
and $f$ is the rotation frequency.
For these two loss mechanisms, the changes in the magnetic field come
from the intrinsic inhomogeneity in the ring magnet, which ideally is
zero.
In practice, however, we measured variations in the magnetic field
around the ring on the order of a few percent.
This inhomogeneity is likely due to manufacturing tolerances.
Note that Equation~\ref{eq:hysteresis} does not depend on the rotation
frequency, while Equation~\ref{eq:eddy} has a linear dependence.
The proportionality constants are ultimately free parameters in the
fit.

The third loss mechanism is produced by the 24 disk magnets used to
move the rotor.
In general, a magnet moving with constant speed above an infinite
conducting plate, will experience magnetic lift and drag forces from
the eddy currents induced in the plate.
Reitz~\cite{reitz_1970} demonstrates that the drag force is
\begin{equation}
F_D= \frac{3 n \mu_0 m^2}{32 \pi z_0^4} \frac{w}{v} \left( 1 -
\frac{w}{\sqrt{ v^2 + w^2 }} \right)
\label{eq:drag}
\end{equation}
where $n$ is the number of magnets, $v$ is the speed, $m$ is the
magnetic dipole moment, $z_0$ is the height of the magnet above the
plate, $w = 2/\mu_0 T \sigma$, and $T$ is the thickness of the plate.
Since the magnets are moving in a circle, $v = 2 \pi f R_{disk}$,
where we assume $R_{disk}$ is the radial distance between the center
of the rotor and the center of each disk magnet (see
Figure~\ref{fig:hrm_and_rotor}).

When taking into account all three loss mechanisms, the equation of
motion for the rotor becomes
\begin{equation}
I \frac{d\omega}{dt} = - R_{disks} \, F_D - R_{ring} \, F_{H} -
R_{ring} \, F_{EC}
\label{eq:eqmotion1}
\end{equation}
where $I = 8.5 \times 10^{-3}$~kg~m$^2$ is the moment of inertia of
the rotor, and we assume $R_{ring}$ is the mean radius of the ring
magnet.
For our fitting procedure, we are interested in determining $f$ as a
function of time.
Therefore, we rearranged Equation~\ref{eq:eqmotion1} to form the more
relevant equation of motion:
\begin{equation}
\frac{df}{dt} = A + B \, f + C F_D(f),
\label{eq:eqmotion2}
\end{equation}
where $F_D(f)$ is simply Equation~\ref{eq:drag} with $v = 2 \pi f
R_{disk}$, and $A$, $B$ and $C$ are the free parameters in the fit.
Note that the unspecified proportionality constants in
Equations~\ref{eq:hysteresis}~\&~\ref{eq:eddy} are accounted for in
the $A$ and $B$ fit parameters, respectively.

We used numerical methods to both solve the differential equation and
fit the solution to the data in Figure~\ref{fig:spin_down}.
The red line corresponds to the best fit model.
Given the values of $A$, $B$, and $C$ from the fit, the gray lines
show how the rotor would spin down if only that loss mechanism
existed.
For example, if we only had hysteresis loss, the rotor would have
slowed down an imperceptible amount in 35~seconds.
The dominant loss mechanism is disk-magnet drag.

Since the rotational kinetic energy in the rotor is $I \omega^2/2$,
during spin down, the energy flowing out of the rotor, in watts, is
\begin{equation}
P = \omega \, I \, \frac{d\omega}{dt},
\label{eq:power}
\end{equation}
which is simply $\omega$ times Equation~\ref{eq:eqmotion1}.
During normal operation we keep the rotation frequency constant with
the drive system, so the equation of motion becomes
\begin{equation}
I \frac{d\omega}{dt} = 0 = \tau_{drive} - \tau_{loss},
\end{equation}
where $\tau_{drive}$ is the applied torque and $-\tau_{loss}$ is the
right hand side of Equation~\ref{eq:eqmotion1}.
This indicates that the mechanical power input by the drive system at
a given rotation frequency is equal to the power ultimately dissipated
into the cryogenic system via $\tau_{loss}$.
It is desirable to minimize the loading on the cryogenics, so
$\tau_{loss}$ should be minimized.
The fit parameters from the spin-down test can be used to calculate
$\tau_{loss}$ for our system, so we can infer the associated heat load
on the cryogenics from this measurement.

\begin{figure}[t]
\centering
\includegraphics[width=\columnwidth]{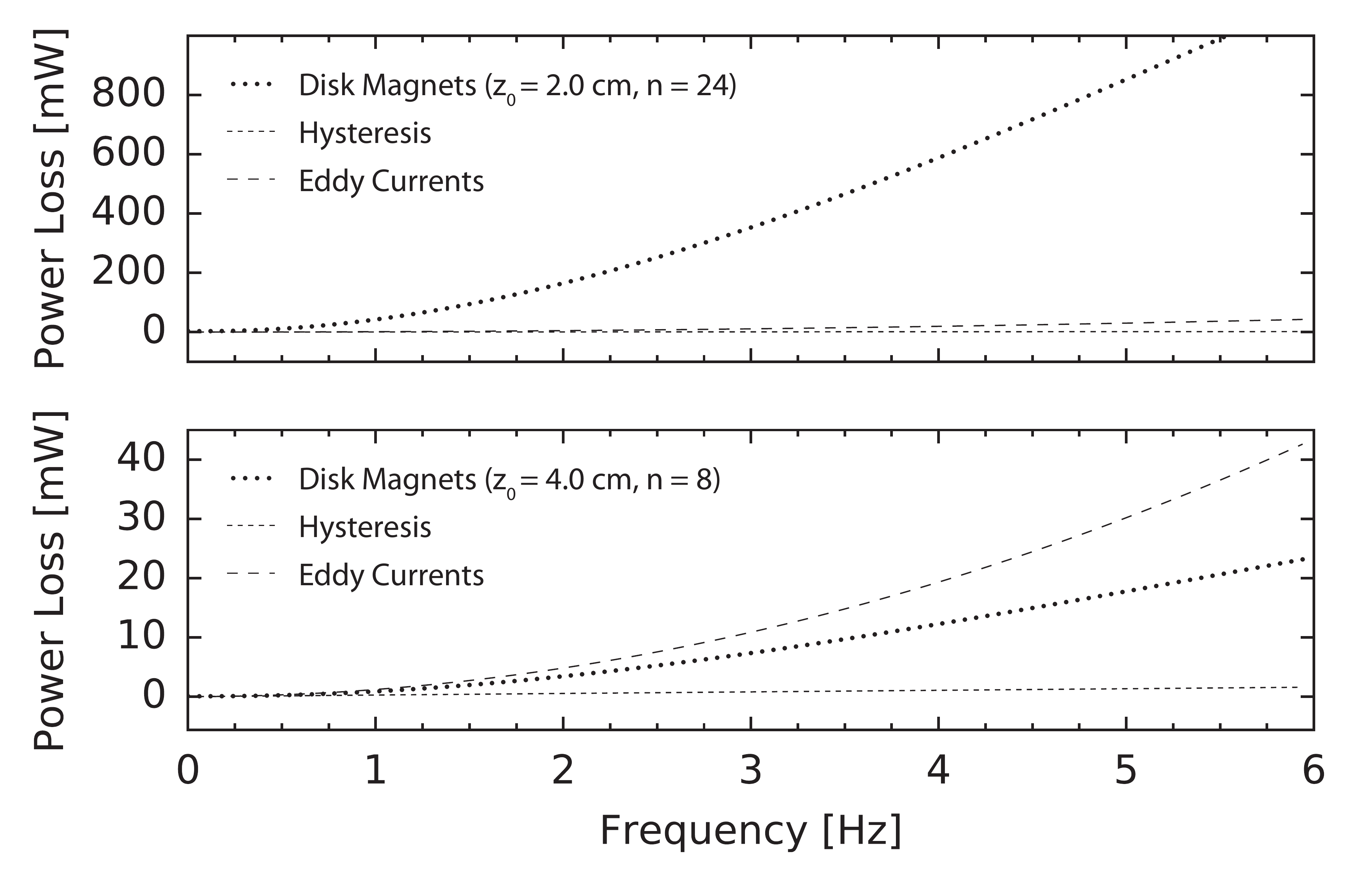}
\caption{
Calculated power loss as a function of rotation frequency.
Points, dotted lines, and dashed lines correspond to contributions
from disk-magnet drag loss, hysteresis loss and eddy current loss,
respectively.
The top panel shows the loss for our configuration ($z_0 =
2.0~\mbox{cm},~n = 24$), and the bottom panel shows the loss for a
modified configuration ($z_0 = 4.0~\mbox{cm},~n = 8$) that is designed
to reduce the amount of loss from disk-magnet drag.
}
\label{fig:power_loss} 
\end{figure}

The top panel of Figure~\ref{fig:power_loss} shows the calculated
power loss for each individual loss mechanism in $\tau_{loss}$ as a
function of rotation frequency.
Power loss from the 24 disk magnets dominates.
Looking at the prefactor in Equation~\ref{eq:drag}, this loss can be
appreciably reduced by increasing the distance between the disk
magnets and the cold plate and decreasing the number of disk magnets
on the torque disk.
In the bottom panel of Figure~\ref{fig:power_loss}, we show how the
calculation result changes if the number of disk magnets is decreased
from from 24 to 8, and the distance between the disk magnets and the
cold plate is increased from 2~cm to 4~cm.
Disk-magnet drag loss decreases by a factor of 48 in this example
configuration, and contributes less than the eddy current loss in the
SMB itself.


\section{Discussion}
\label{sec:discussion}

Our long-term goal is to develop a hollow-shaft, SMB-based motor with
a $\sim$500~mm aperture that operates at 3~K for next-generation
experiments\cite{cmb-s4_technology_book, cmb-s4_science_book}.
We built and tested the prototype motor presented in this paper to
demonstrate the new and more risky subsystems that are needed for the
more ambitious large motor.
In particular, we were most interested in demonstrating (i) the
viability of the scalable HRM design, (ii) the encoder readout
approach, and (iii) the idea that the encoder signals can be used in
the feedback loop.
The prototype motor test was successful.

Before building the larger motor, several issues need to be addressed.
First, the loss from the tested prototype configuration is
significant, and it needs to be decreased in future designs.
The precise loss requirement will likely depend on the operating
temperature choice and whether the future cryogenic system is
mechanical-cooler-based or liquid-cryogen-based.
Nevertheless, as mentioned in Section~\ref{sec:loss}, the drag-force
loss from the disk magnets can be decreased below the eddy current
loss in the SMB with straightforward design changes.
This target seems reasonable because decreasing the eddy current loss
in the SMB will be challenging, so it serves as a fundamental limit.
In addition, it may also be worth making the metal structures on the
stator out of a superconductor with a transition temperature greater
than 3~K, so $\sigma \rightarrow \infty$.
This change, in principle, makes the disk-magnet drag force zero (see
Equation~\ref{eq:drag}).
Also, the steel screws that hold the ring magnet in the rotor could
introduce a small azimuthal pattern in the magnetic field of the ring
magnet ($\Delta B$ in Equations~\ref{eq:hysteresis}~\&~\ref{eq:eddy}),
which could be a source of the energy loss.
This effect should be examined more carefully.
Second, one source of heat we did not consider in this study is Joule
heating in the coil wire, which can also be appreciable.
Making the coils out of superconducting niobium-titanium wire should
eliminate this heat source at 3~K.
Third, millimeter-wave detector systems can be sensitive to magnetic
fields.
Therefore, the scaled-up motor should include flux returns and a
$\mu$-metal enclosure to minimize stray magnetic fields.
Fourth, the rotor temperature, or more importantly, the HWP
temperature, should be measured during operation.
This measurement is difficult because the rotor is levitating.
Therefore, a wireless temperature sensor needs to be developed.
Fifth, rotor vibration needs to be considered~\cite{sakurai_2017a}.
Finally, the rotor temperature should be as close to the 3~K stator
temperature as possible.
Some of the light in the telescope beam will be absorbed by the HWP,
which introduces a heat load on the levitating rotor.
This heat must me removed through radiation, so a radiative heat
exchanger between the rotor and the stator needs to be developed.

For the larger motor, it should be possible to decrease the
orientation angle uncertainty.
For example, if the slit pitch remains constant (2.22~mm) and the
encoder disk diameter increases from 254~mm to 508~mm, then there will
be a total of 720 slits instead of 360 slits.
Therefore, by simply counting slits with the optical fiber encoder
sensors, the upper limit on the orientation angle uncertainty
decreases from 0.5~deg to 0.25~deg, which is very close to our
requirement.
This uncertainty can be further decreased, as desired, by increasing
the total number of slits, which can be achieved by decreasing the
slit pitch further, increasing the encoder disk diameter, or both.
 

\begin{acknowledgments}

We would like to thank Robert Rickenbach at Micronor for his
assistance and design input regarding the encoder.
We would also like to thank Daniel Flanigan, Mark Greenan, David
Colavita, and Joshua Sobrin for their help.
This project was supported in part by an Italian Space Agency
fellowship for Columbro and an NSF Astronomy and Astrophysics
Postdoctoral Fellowship for Reichborn-Kjennerud.

\end{acknowledgments}


\bibliography{bibliography} 


\end{document}